\documentclass[lettersize,journal]{IEEEtran}
\usepackage{cite}

\usepackage[spanish, english]{babel}

\usepackage{float}
\usepackage{caption}
\usepackage{hyperref}
\usepackage{xcolor}
\usepackage{enumerate} 

\usepackage{amsmath,amsfonts,stmaryrd,amssymb, amsthm}
\usepackage{mathtools}
\usepackage{bbm}

\newcommand{\trans}[1]{\strut{#1}^{{\mkern-2mu\mathsf{T}}}}

\newcommand{\diff}{\mathop{}\!\mathrm{d}}

\newtheorem{proposition}{Proposition}

\def\etal{\emph{et al.~}}

\newcommand{\bt}[1]{\mbox{$\bf #1$}}
\def\l{\left(}
\def\r{\right)}

\newcommand{\x}{\bt x}
\newcommand{\res}{\bt w}

\newcommand{\nimag}{L}
\newcommand{\npix}{N}

\newcommand{\pow}{\sigma_k^2}
\newcommand{\powsq}{\sigma_k}
\newcommand{\indexv}[2]{[#1]_{#2}}

\newcommand{\bigo}[1]{\mathcal{O}(#1)}

\newcommand{\newnoise}{\bt n}
\begin{document}
	
	\title{{PRNU Emphasis: a Generalization of the Multiplicative Model}}
	
	\author{Samuel~Fernández-Menduiña,
		Fernando~Pérez-González,~\IEEEmembership{Fellow,~IEEE},
		{and~Miguel~Masciopinto}
		\thanks{
			 This work was funded by Agencia Estatal de Investigacion (Spain) and the European Regional Development Fund (FEDER ”Una manera de hacer Europa”) under project RODIN (PID2019-105717RB-C21), and by Xunta de Galicia "Grupos de Referencia Competitiva" (ED431C 2021/47).
		}
		\thanks{
			The authors are with atlanTTic, University of Vigo,  Department of Signal Theory and Communications, 36310 Vigo, Spain. (e-mail: sfmenduina@gts.uvigo.es; fperez@gts.uvigo.es; mmasciopinto@gts.uvigo.es).}
	}
	
	
	\IEEEpubid{}
	
	\maketitle
	
	\begin{abstract}
		The photoresponse non-uniformity (PRNU) is a camera-specific pattern, widely adopted to solve multimedia forensics problems such as device identification or forgery detection. The theoretical analysis of this fingerprint customarily relies on a multiplicative model for the denoising residuals. This setup assumes that the nonlinear mapping from the scene irradiance to the preprocessed luminance, that is, the composition of the Camera Response Function (CRF) with the optical and digital preprocessing pipelines, is a gamma correction. Yet, this assumption seldom holds in practice. In this letter, we improve the multiplicative model by including the influence of this nonlinear mapping on the denoising residuals. We also propose a method to estimate this effect. Results evidence that the response of typical cameras deviates from a gamma correction. Experimental device identification with our model increases the TPR by a $4.93\, \%$ on average for a fixed FPR of $0.01$.
	\end{abstract}
	
	\begin{IEEEkeywords}
		PRNU, Camera response function, Sensor fingerprint, Sensor pattern noise, Device identification, Image forensics.
	\end{IEEEkeywords}
	
	\section{Introduction}
	
		The photoresponse non-uniformity (PRNU) is a noise-like pattern due to manufacturing imperfections. Being sensor intrinsic, it qualifies as a device fingerprint for multimedia forensics problems, including forgery detection and device identification \cite{ChenFGL2008}. Since the pattern is buried under the image content, signal processing plays a key role in handling the PRNU \cite{LukasFG2006}. In this sense, although data-driven algorithms are promising \cite{KirchnerJ2019, CozzolinoV2019}, model-driven strategies still prevail. 
				
		In the research literature, the multiplicative model proposed by Chen \etal \cite{ChenFGL2007} is well-established. Following this setup, the PRNU can be estimated from a set of denoising residuals using the estimator proposed in \cite{PerezMIC2016}. Through a hypothesis testing problem, this estimate is compared with the denoising residual of a test image, often computing the PCE \cite{GoljanF2008}. When the number of pixels is large, the multiplicative model leads to almost optimal PRNU estimators and detectors \cite{PerezMIC2016}.

		Recently, the multiplicative model has been brought into question \cite{MasciopintoP2018}. Using synthetic signals, Masciopinto and Pérez-González showed that some of the predictions based on the multiplicative model do not conform with the empirical results. Motivated by these findings, in this letter, we put forward a new model based on the theoretical analysis of the transfer curve, a term we use to denote the composition of the camera response function (CRF) with the optical and digital preprocessing pipelines.	
			
		The CRF is the mapping from image irradiance to image brightness. Evidence suggests that the response function ``can vary significantly from a gamma curve'' \cite{GrossbergN2004}. Moreover, in practice, the optical and digital preprocessing pipelines, including the ISO and the demosaicing, affect the captured image. Yet, as we will show, the multiplicative model is only valid when this transfer curve is a gamma correction.
			
		Thus, in this letter, we show how to derive a generalized model accounting for this composition of functions. This generalization includes the multiplicative model presented in \cite{ChenFGL2007} as a special case (Sec.~\ref{sec:genmod}). The resulting expression shows that emphasizing some brightness values of the denoised image when handling the fingerprint has benefits. Hence, we call this effect \emph{PRNU emphasis}; we also propose a method to estimate it (Sec.~\ref{sec:rec_emph}).
		
		We note that a related idea was proposed by Li under the name ``SPN enhancers'' \cite{Li2010}. However, we depart from a model for the captured image, while SPN enhancers were derived from a hypothesis about the relation between the denoising residual and the scene content. The result is also different: SPN enhancers act on the denoising residual, while PRNU emphasis acts on the denoised image.
		
		To validate our proposal, we estimated the PRNU emphasis from real images. The results differ from the linear function assumed by the multiplicative model. Then, we address a device identification problem with cropped images using the estimated PRNU emphasis. We also test a fixed emphasis function, which requires less computations. Our model yields a higher true positive rate (TPR) than the multiplicative model (Sec.~\ref{sec:exper}).
		
		\textbf{Notation.} Uppercase bold letters, such as $\bt A$ or $\bt \Phi$, denote matrices. Lowercase bold letters, such as $\bt a$, denote vectors. We denote the $n$th component of the vector $\bt a$ by $\indexv{\bt a}{n}$. We use $\bt 1$ and $\bt 0$ to denote the vector of all-ones and all-zeros, respectively. Regular letters, either Latin or Greek, denote scalar values.
		\section{{Generalized} PRNU Model}
		\label{sec:genmod}
		Let us recall the multiplicative model\cite{ChenFGL2007}: Applying a first-order Taylor expansion to the output of the gamma correction with parameter $\gamma$, we can write, for a set of $L$ images,
		\begin{equation}
			{\bt y_l = \bt x_l \circ ( \bt 1 + \gamma \, {(\powsq \bt k))} + \bt n_l}, \quad l = 0, 1, \hdots, L-1,
		\end{equation} 
		where $\bt y_l$ is the sensor output, $\x_l$ denotes the gamma-corrected noiseless image, and $\bt k$ denotes the unit variance PRNU, all of them $\npix$-dimensional vectors obtained by vectorizing the original images of size $N_1\times N_2 = N$. The symbol $\circ$ denotes the entry-wise product. The variance of the PRNU is $\pow$. 
		
		We assume the fingerprint $\bt k$ follows a zero-mean white Gaussian process with unit variance. Applying a denoising filter, we reach
		\begin{equation}
			\label{eq:chenetal}	
			\res_l = {\gamma}\, \x_l \circ (\powsq \bt k) + \bt n_l, \quad l = 0, 1, \hdots, \nimag-1.
		\end{equation}
		Although denoisers are not ideal \cite{FernandezP2021, PerezF2021}, we assume $\bt n_l$ are uncorrelated processes, which are also uncorrelated with $\bt k$.
		\subsection{{Transfer curve}}
		\label{sec:emph_th}
\begin{figure}[t]
		\centering
		\includegraphics[scale = 0.27]{"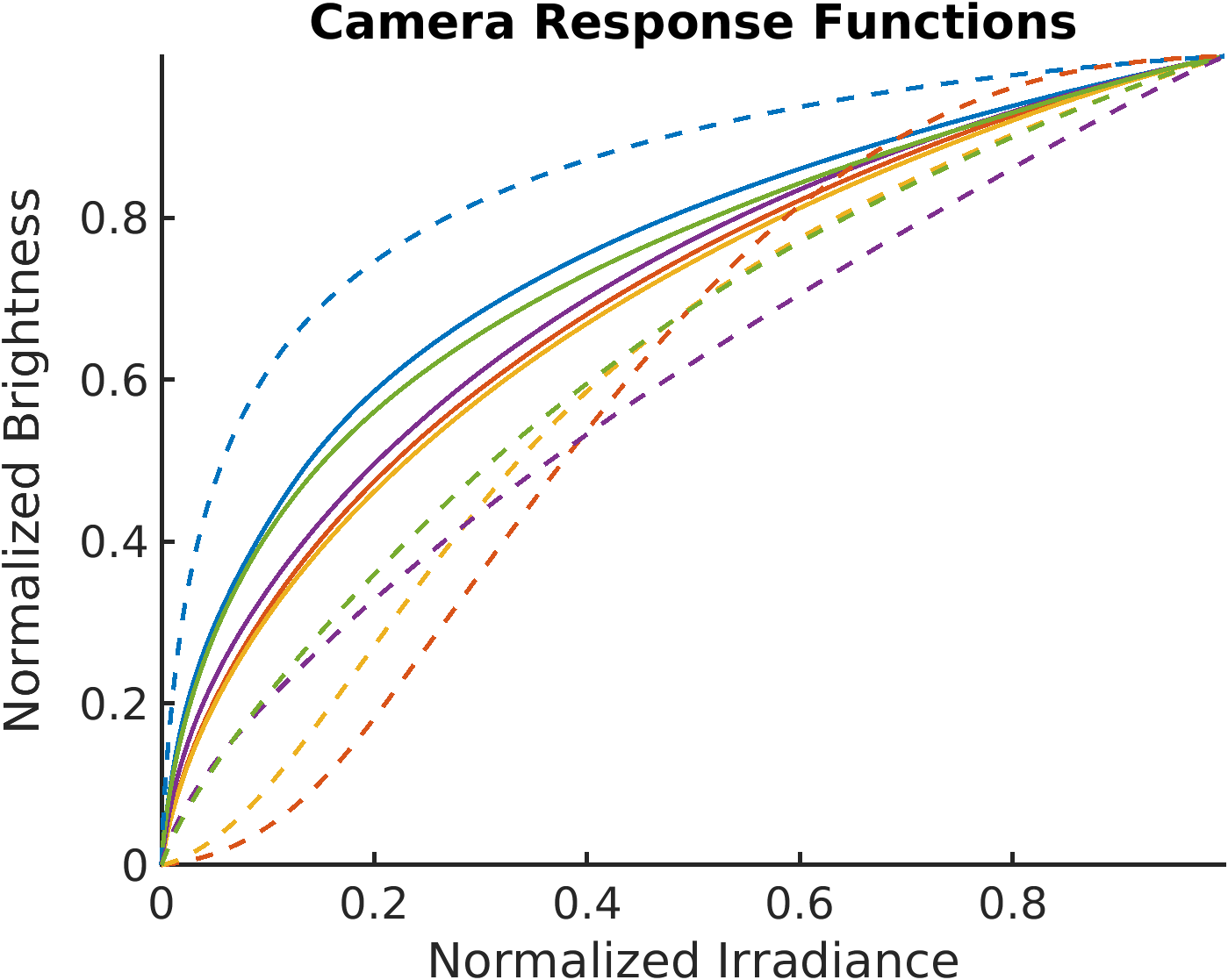"} \hfill
		\caption{Ten CRFs of real cameras, from \cite{DoRF}.}
		\label{fig:cam_res_fun}
	\end{figure}

		The gamma correction is a Camera Response Function (CRF), or the mapping from image irradiance to image brightness \cite{GrossbergN2004}. This mapping is a monotonically increasing function between $0$ and $1$ \cite{GrossbergN2004}. In Fig.~\ref{fig:cam_res_fun}, we depict some examples of CRFs: they differ from a gamma curve.
		
		Moreover, we need to account for the optical and digital preprocessing pipelines. We call this composition of functions the transfer curve, $h$. We assume it is monotonically increasing. Two conditions that ensure this property are 1) operations applied pixel-wise, like the ISO, are well approximated by monotonic functions, and 2) operations that mix different pixels, such as demosaicing \cite{GunturkGASM2005} or gray-scale conversion, are well approximated by a positive linear combination of these pixels. In Sec. \ref{sec:exper}, we test this assumption empirically. We also set $h(1) = 1$ and $h(0) = 0$ \cite{GrossbergN2004}. Let $\bt z$ be the input to the imaging sensor. Then, the output of the sensor is:
		\begin{equation}
			\label{eq:grossnay}
			\bt y  = h\l \bt z + \bt z \circ {(\powsq \bt k)} \r + \newnoise.
		\end{equation} 
		The function $h$ acts element-wise: $\indexv{h(\bt a)}{n} = h(\indexv{\bt a}{n})$ for $n = 0, \hdots, \npix-1$.  Our goal is to show that the transfer curve influences the denoising residuals. First, we assume $h$ is differentiable in its domain; let ${h'(x)}$ be the derivative.  Applying the first-order Taylor expansion of {$h\l \bt z + \bt z \circ (\powsq \bt k)\r$} about the point $\powsq\bt k = \bt 0$, we obtain
		\begin{equation}
			\bt y =  h(\bt z) + \bt z \circ h'(\bt z) \circ {(\powsq \bt k)} + \bigo{\pow \, \bt k^2} + \newnoise.
		\end{equation}
		Now, we set $\bt x \doteq h(\bt z)$. As $\bt k$ has zero mean, the output of the denoiser is $h(\bt z)$ and the residual can be written as
	\begin{equation}
		\label{eq:newmod}
		\res = g(\x) \circ {(\powsq \bt k)} + \bt n_d,
	\end{equation} 
	where $ g(z) \doteq z \, h'(z)$ acts element-wise and $\bt n_d$ collects the noise terms. We call the function $g$ the \emph{PRNU emphasis}; in Sec.~\ref{sec:rec_emph}, we show how to estimate it. This model matches \eqref{eq:chenetal} if and only if $h$ is a gamma correction. See the Appendix.  
	
	\section{Estimating the PRNU Emphasis}
	\label{sec:rec_emph}
	To estimate the PRNU emphasis, we assume the average of the deviations introduced by the denoiser converges to zero.
	
	\subsection{{Covariance}}
	Given the $n$th entry of two residuals, $\indexv{\res_i}{n}$ and $\indexv{\res_j}{n}$, obtained from different images but taken with the same camera, their covariance is $\pow\, \indexv{g(\bt x_i)\circ g(\bt x_j)}{n}$ for $n = 0, 1, \cdots, N-1$. Since we know the vectors $\x_i$ and $\x_j$ from the denoiser, we can define the following bivariate function:
	\begin{equation}
		\label{eq:phi}
		\phi: (z_1, z_2) \mapsto \pow \, g(z_1)g(z_2).
	\end{equation}
	Then, our goal is to estimate $\phi$, which binds the pair $(\indexv{\x_i}{n}, \indexv{\x_j}{n})$ with the covariance between $\indexv{\res_i}{n}$ and $\indexv{\res_j}{n}$.
	
	\subsection{Regressogram}
	\label{sec:reg}
	If the function to estimate is continuous, we can use a regressogram \cite{Tukey1961}. Let $\hat{\phi}$ denote the estimate of the function $\phi$. Our two covariates have the same span. Thus, we divide the $2$D plane in squares. Let $I_{p, q}$ be the square corresponding to the $p$th division in one axis and the $q$th division in the other axis. Then, for $(z_1, z_2)\in I_{p, q}$, the regressogram is
	\begin{equation}
		{\hat{\phi}(z_1, z_2) =} \frac{{\sum_{i, j\not = i}}\sum_n \indexv{\bt w_i \circ \bt w_j}{n}\ {\mathbbm{1}_{p, q}((\indexv{\bt x_i}{n}, \indexv{\bt x_j}{n}))}}{{\sum_{i, j\not =  i}}\sum_n {\mathbbm{1}_{p, q}(( \indexv{\bt x_i}{n}, \indexv{\bt x_j}{n}))}},
	\end{equation}
	where $\mathbbm{1}_{p, q}$ is the indicator function of $I_{p, q}$.
	\begin{figure*}[t]
		\centering
		\includegraphics[scale = 0.22]{"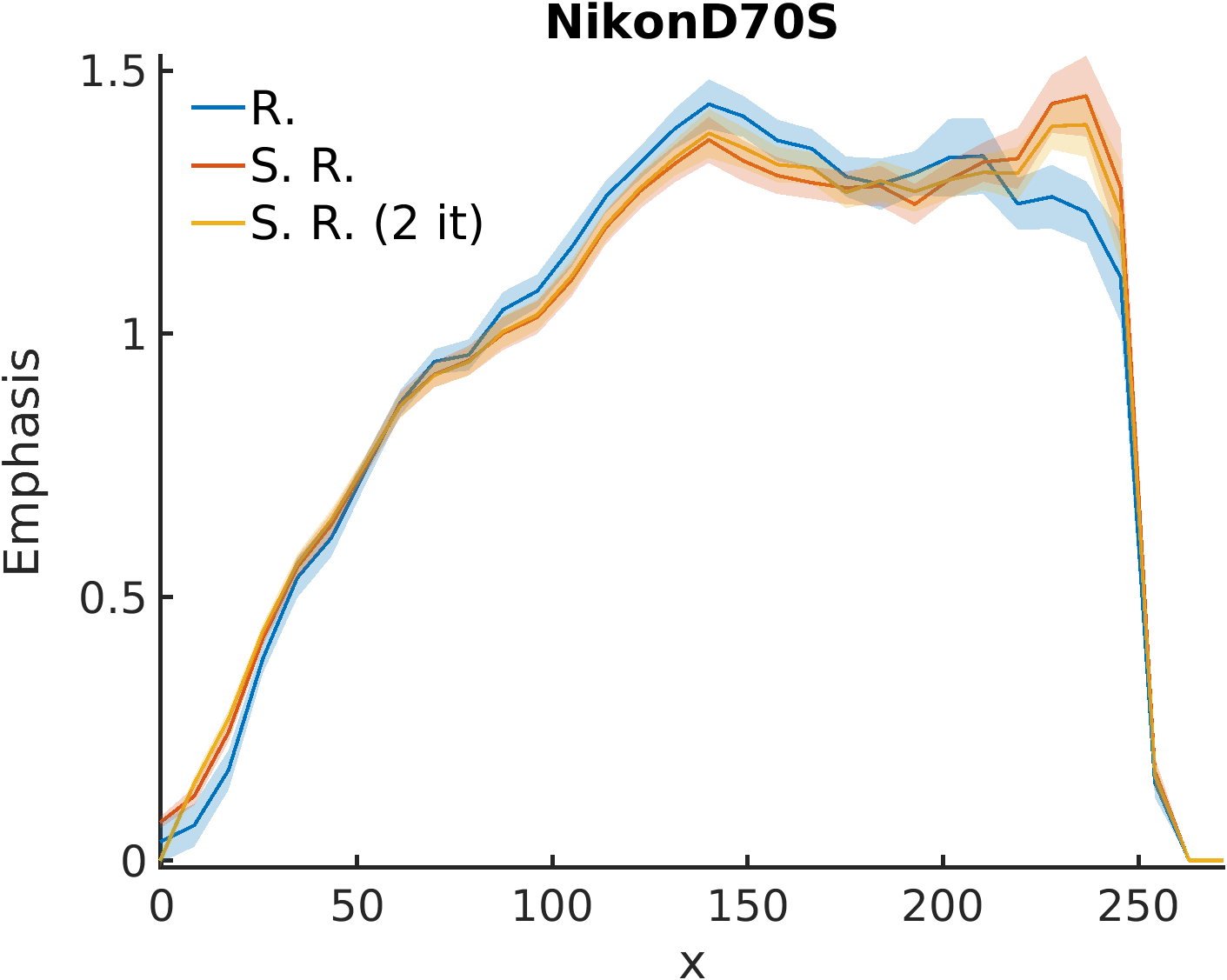"}
		\hspace{2em}
		\includegraphics[scale = 0.22]{"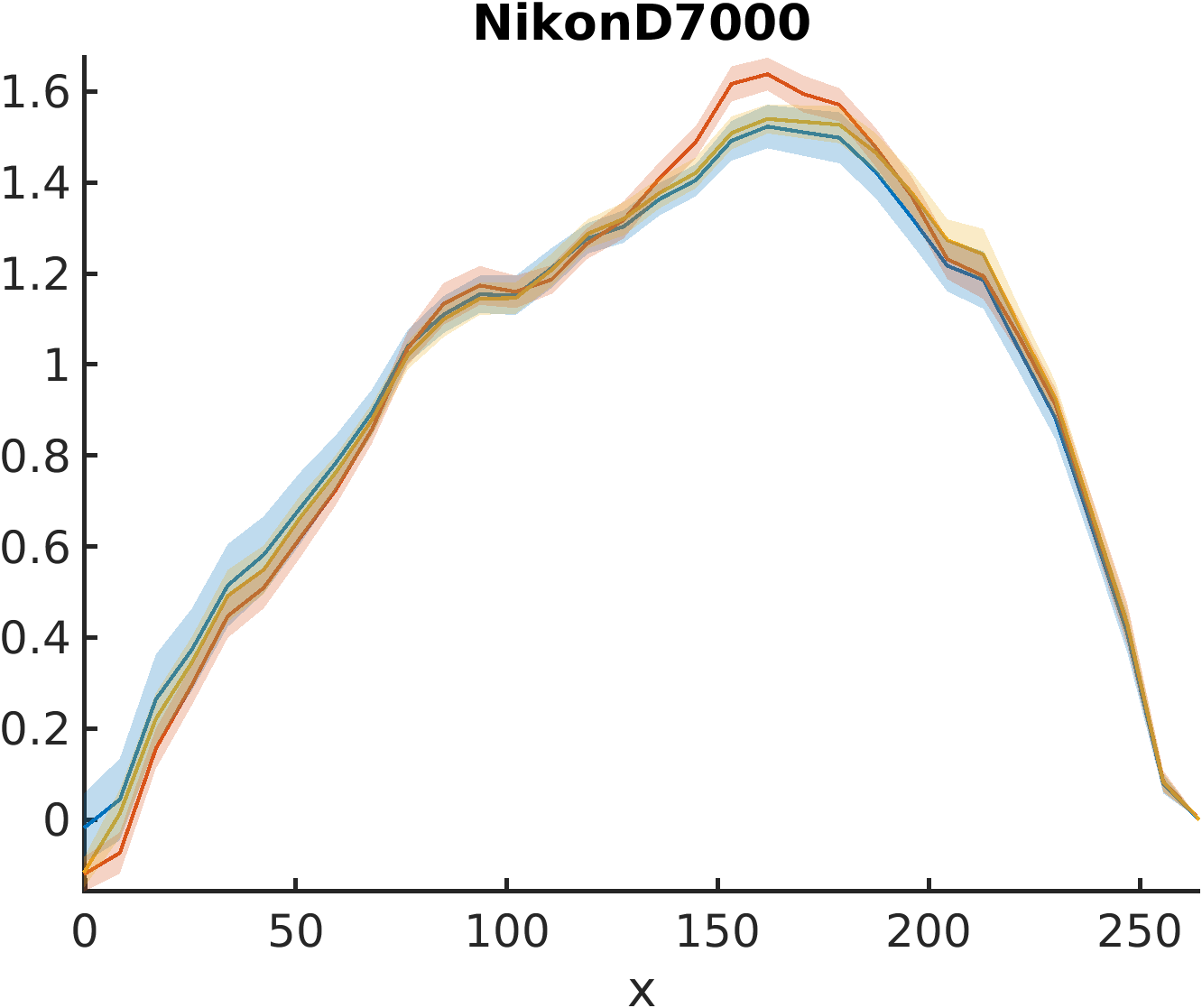"}
		\hspace{2em}	
		\includegraphics[scale = 0.22]{"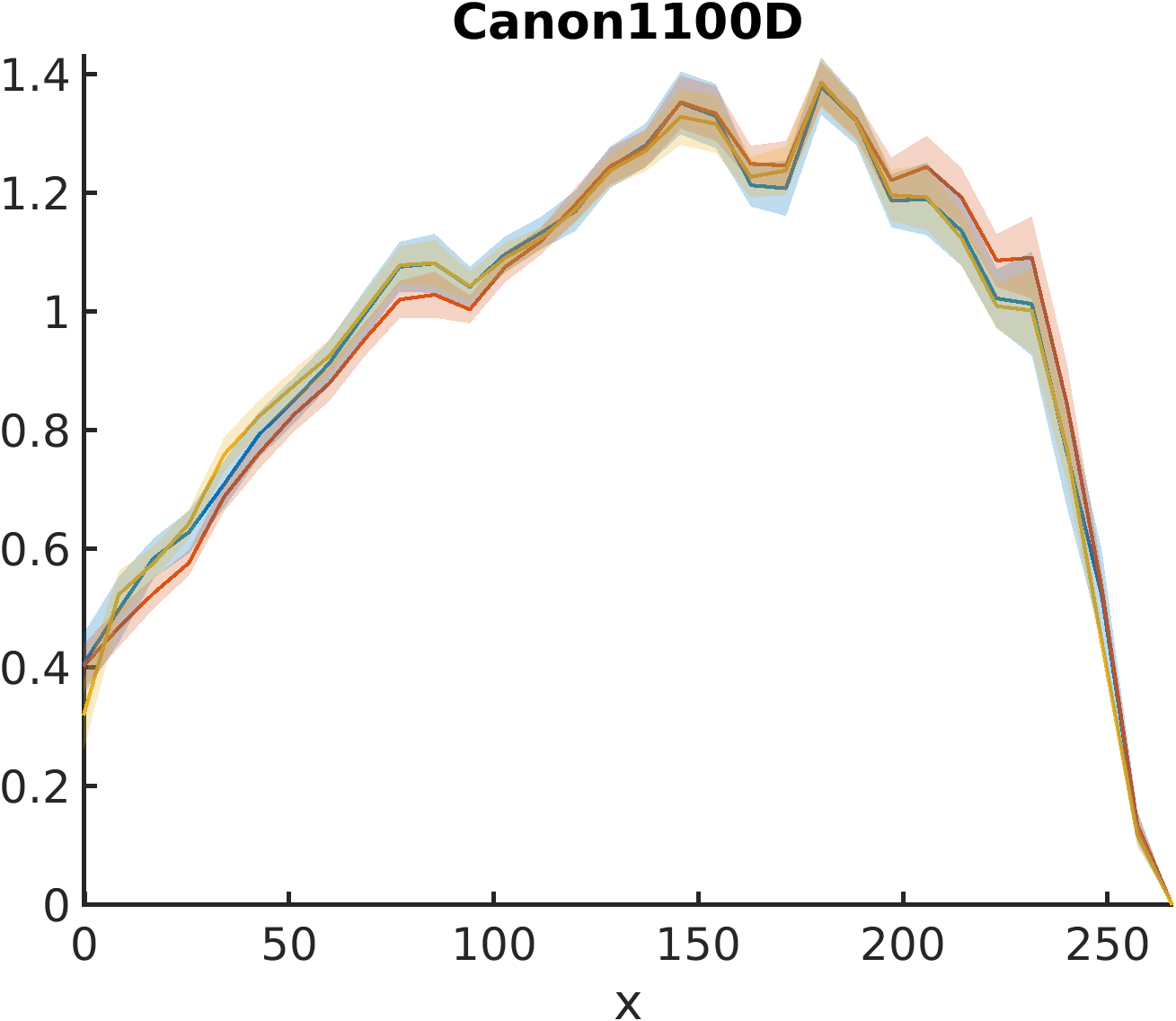"}		
		 \hfill
		
		\vspace{1em}
		
		\centering 

		\includegraphics[scale = 0.22]{"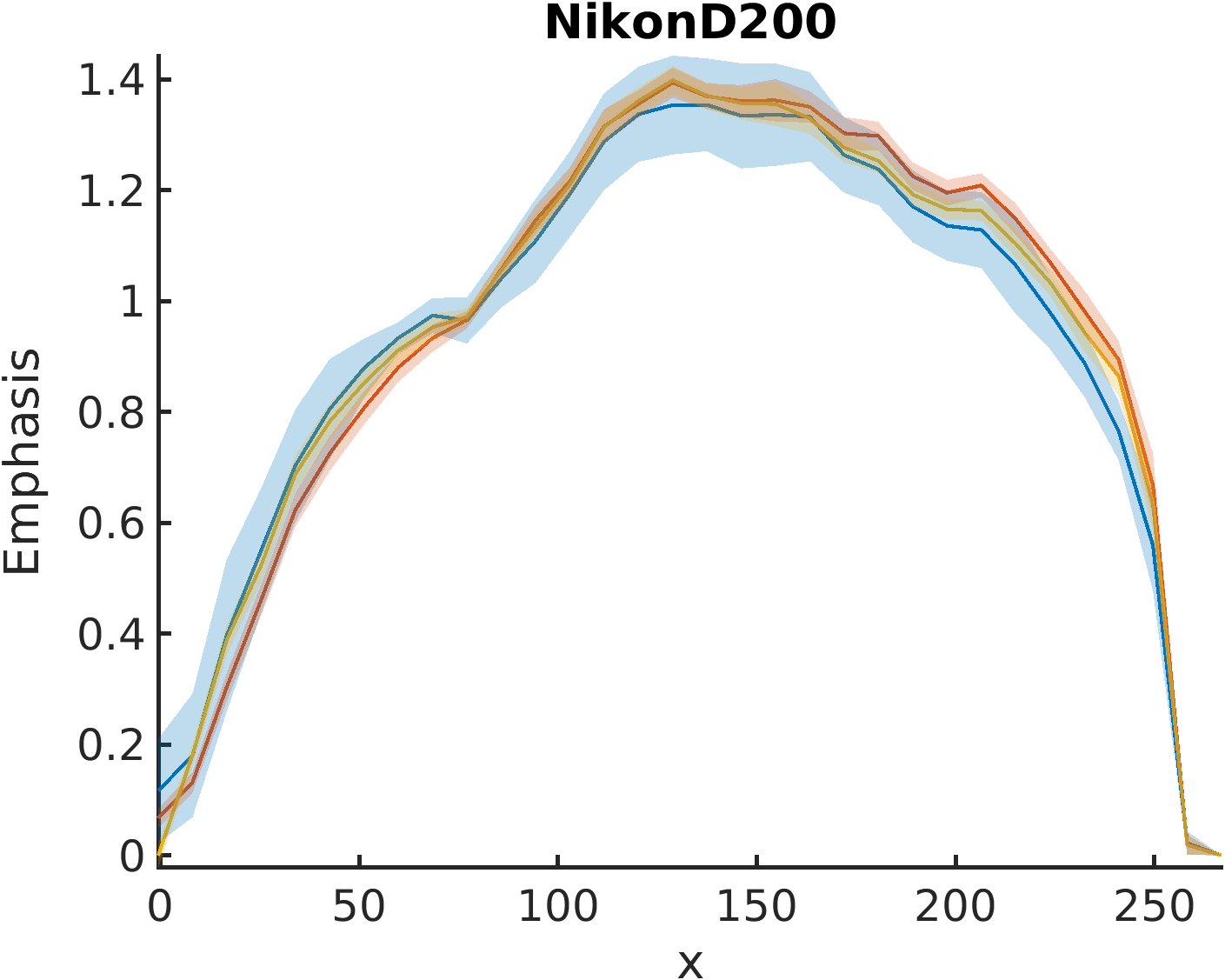"} 
		\hspace{2em}		 		
		\includegraphics[scale = 0.22]{"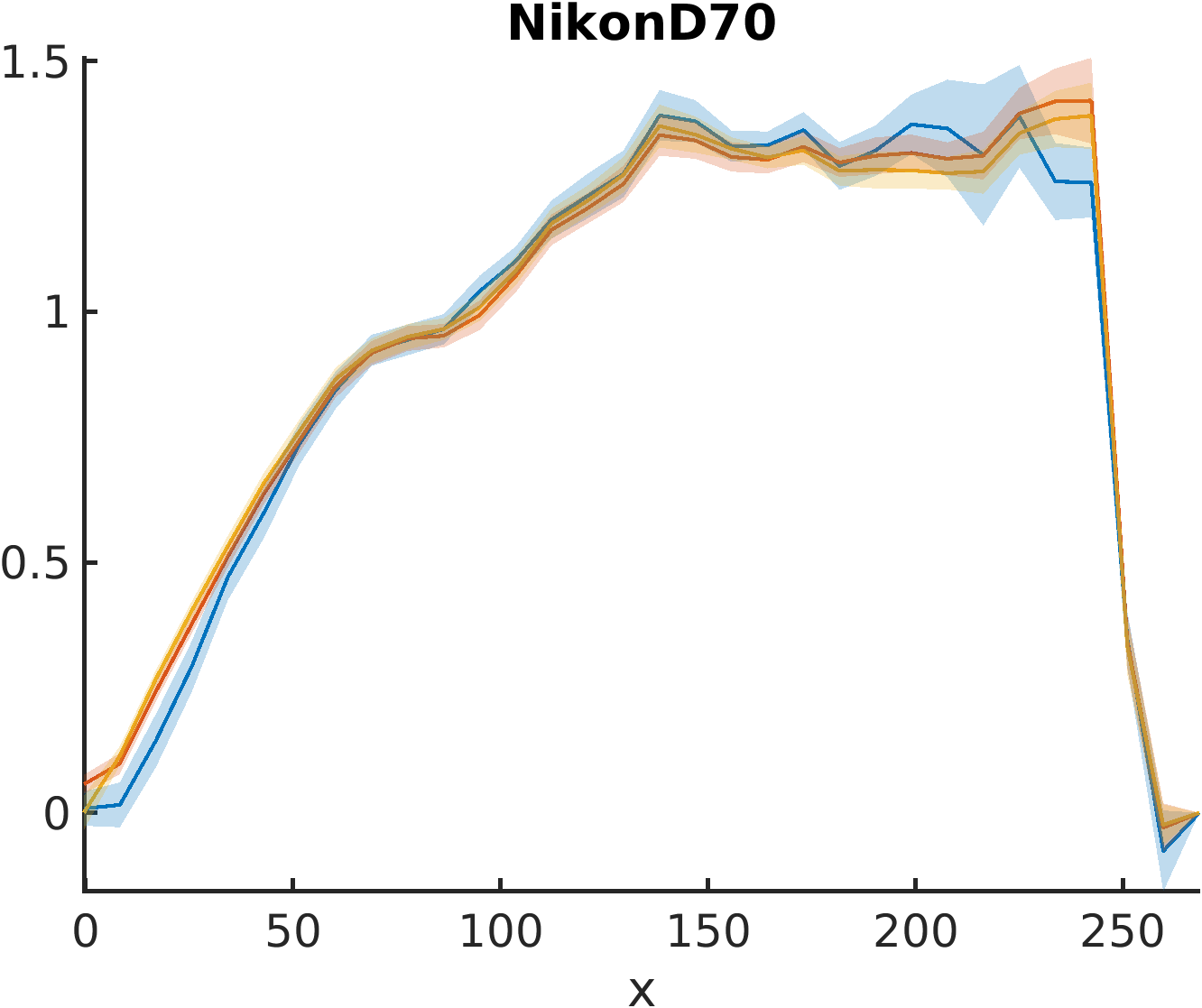"}	
		\hspace{2em}	
		\includegraphics[scale = 0.22]{"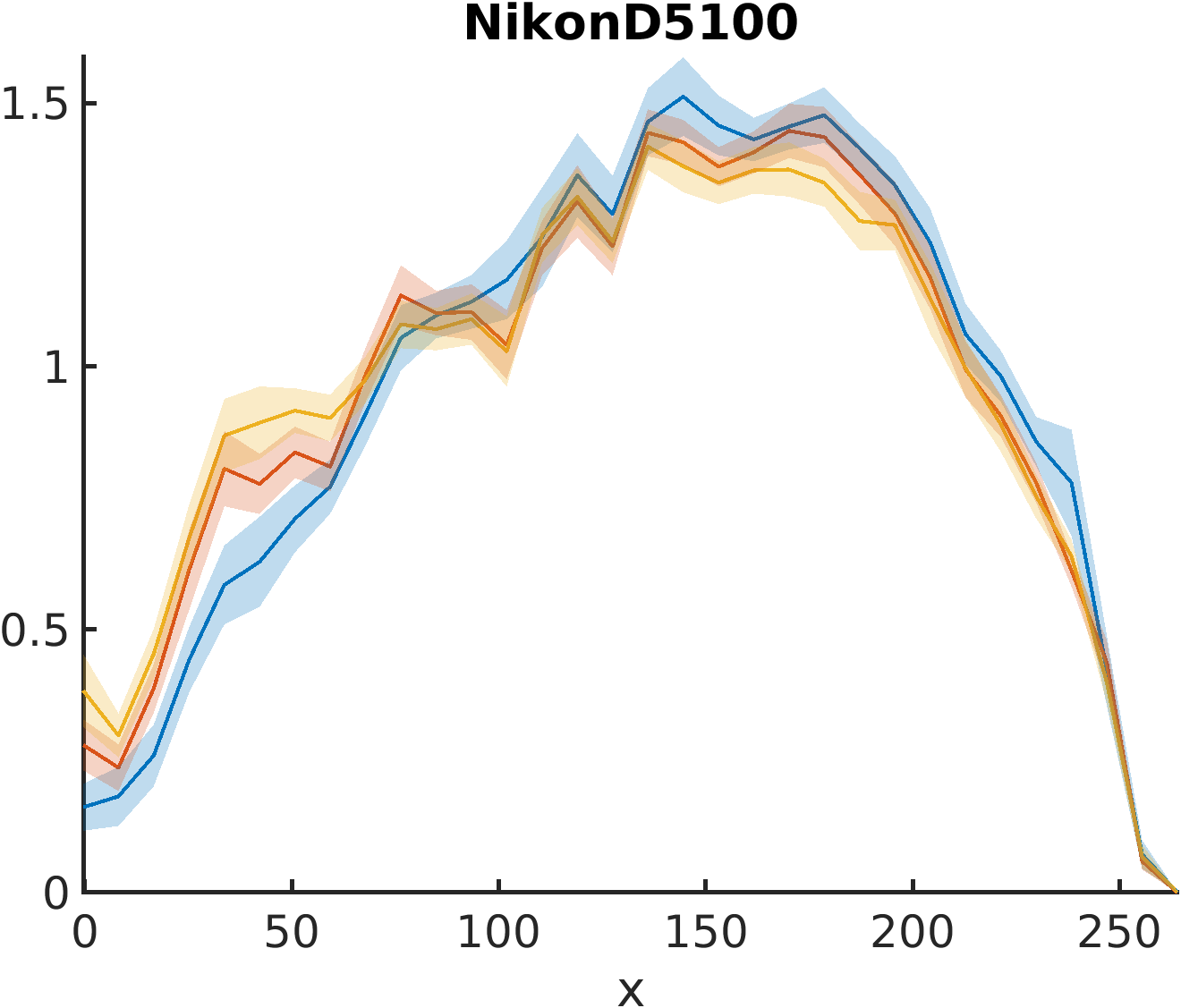"}	 
		 \hfill 	
		\caption{Estimated PRNU emphasis. Blue: regressogram (R); red: simplified regressogram (S. R.); yellow: simplified regressogram, two iterations (S. R. 2 it). The semi-transparent bands are the $95\,\%$ Gaussian confidence intervals.}
		\label{fig:regress_real}
	\end{figure*}
	
	\begin{figure}[t]
		\centering
		\includegraphics[scale = 0.27]{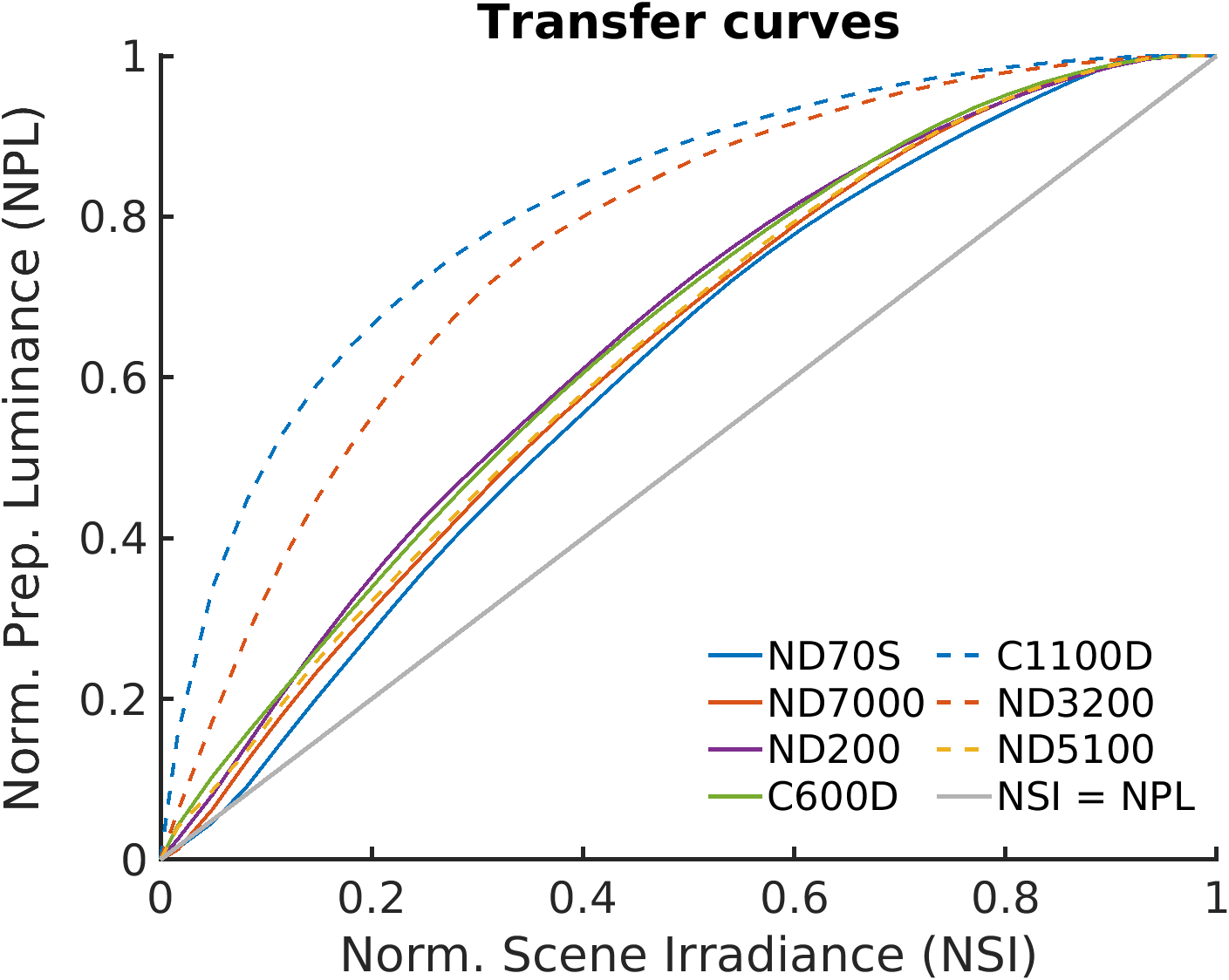} \hfill
		\caption{Estimated transfer curves. ``Prep.'' stands for ``preprocessed'' and ``Norm.'' for ``normalized''.}
		\label{fig:CRFs}
	\end{figure}
	\subsection{From the regressogram to the PRNU emphasis}
	 Let us define the matrix $\hat{\bt \Phi}$, that contains the values of {the regressogram} for each possible $p$ and $q$. This matrix is symmetric because $\phi(z_1, z_2) = \phi(z_2, z_1)$ holds. Also, in the noiseless scenario, $\hat{\bt \Phi}$ has rank one since it is the product of a function by itself sampled at the same points in both dimensions. 
	
	We can exploit these properties to recover $g(x)$ as follows. First, we symmetrize the estimation averaging the matrix and its transpose. We call the result $\hat{\bt \Phi}_s$.  
	Now, by the Eckhart-Young theorem \cite{GolubV:2013}, we know that the closest rank one matrix to $\hat{\bt \Phi}_s$ in the $2$-norm sense is $\sigma_1 \, \bt g \trans{\bt g}$, where $\sigma_1$ is the largest singular value of $\hat{\bt \Phi}_s$ and the vector $\bt g$ is the singular vector associated with $\sigma_1$. This vector $\bt g$ approximates the samples of $g(x)$.
	\begin{figure*}[t]
		\centering
		\includegraphics[scale = 0.18]{"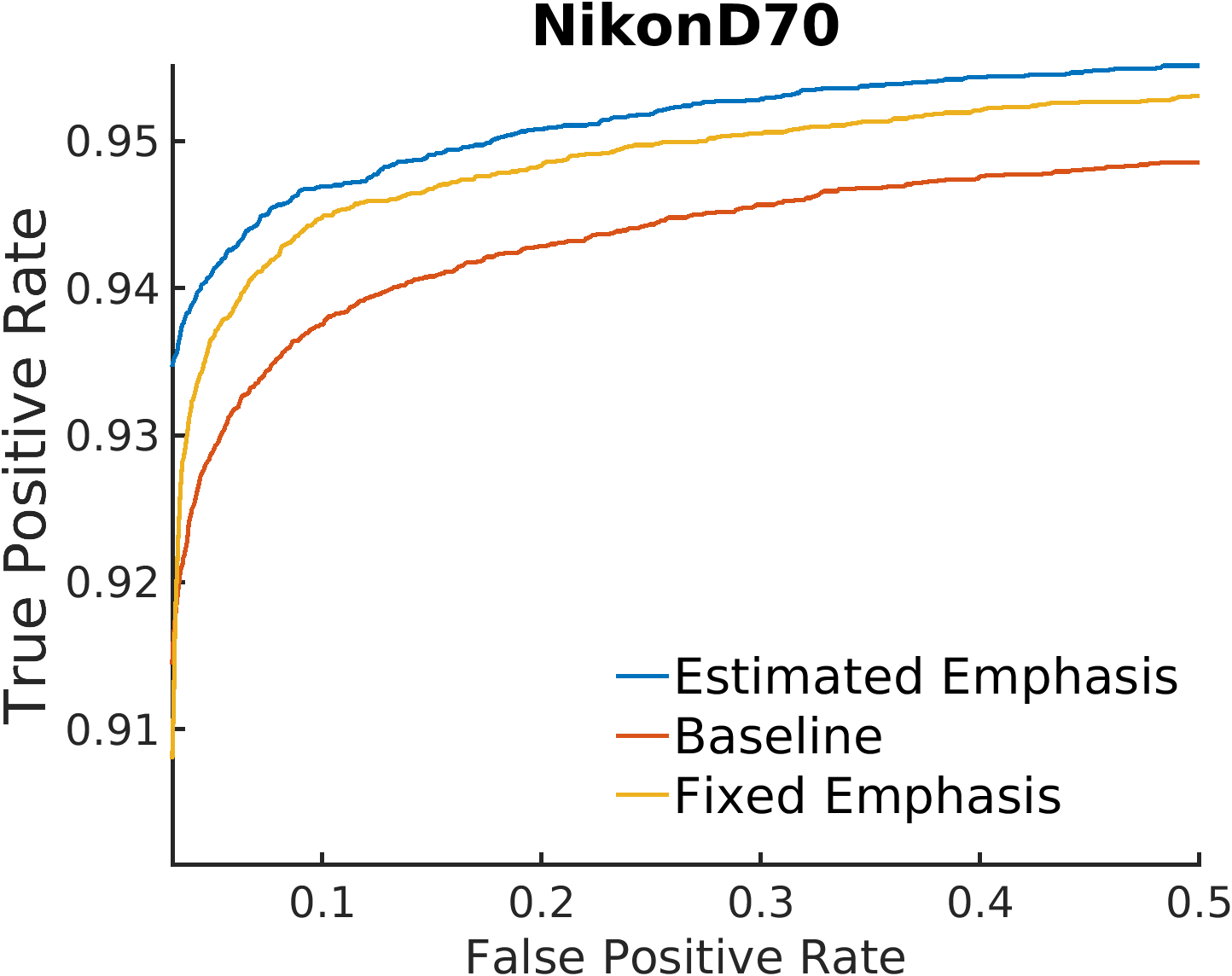"} 
		\hspace{0.01em}	 
		\includegraphics[scale = 0.18]{"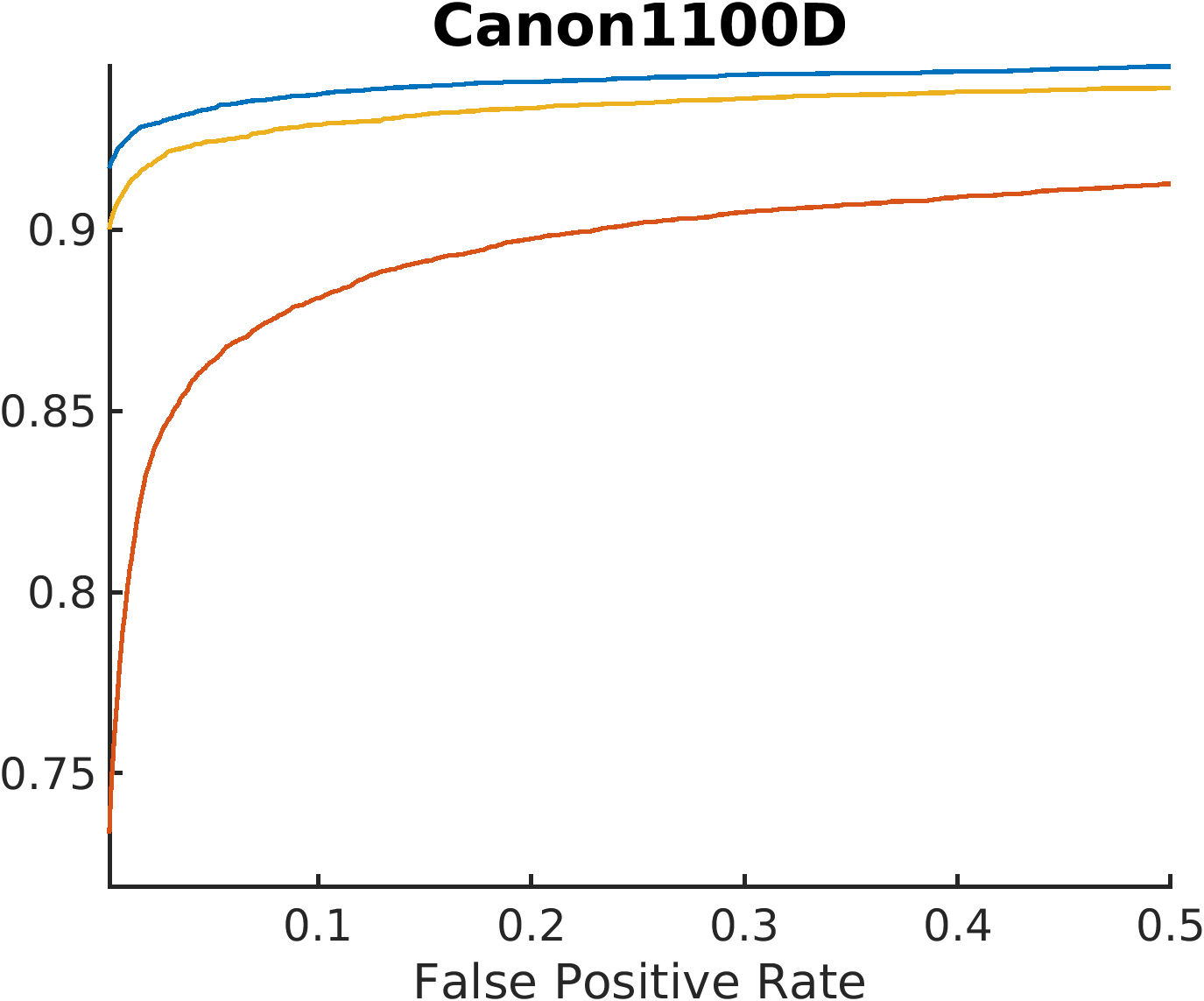"}
		\hspace{0.01em} 
		\includegraphics[scale = 0.18]{"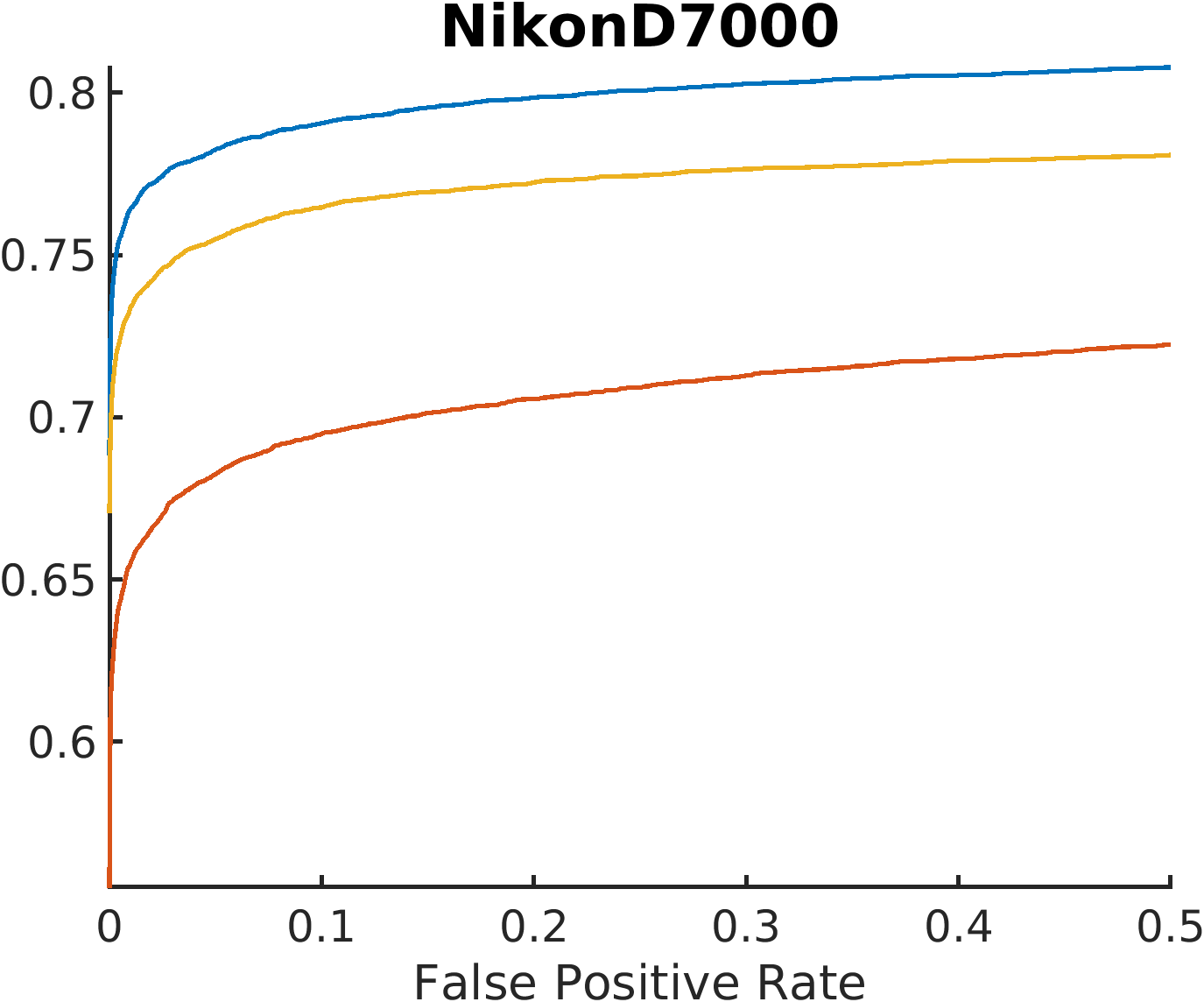"}
		\hspace{0.01em} 
	\includegraphics[scale = 0.18]{"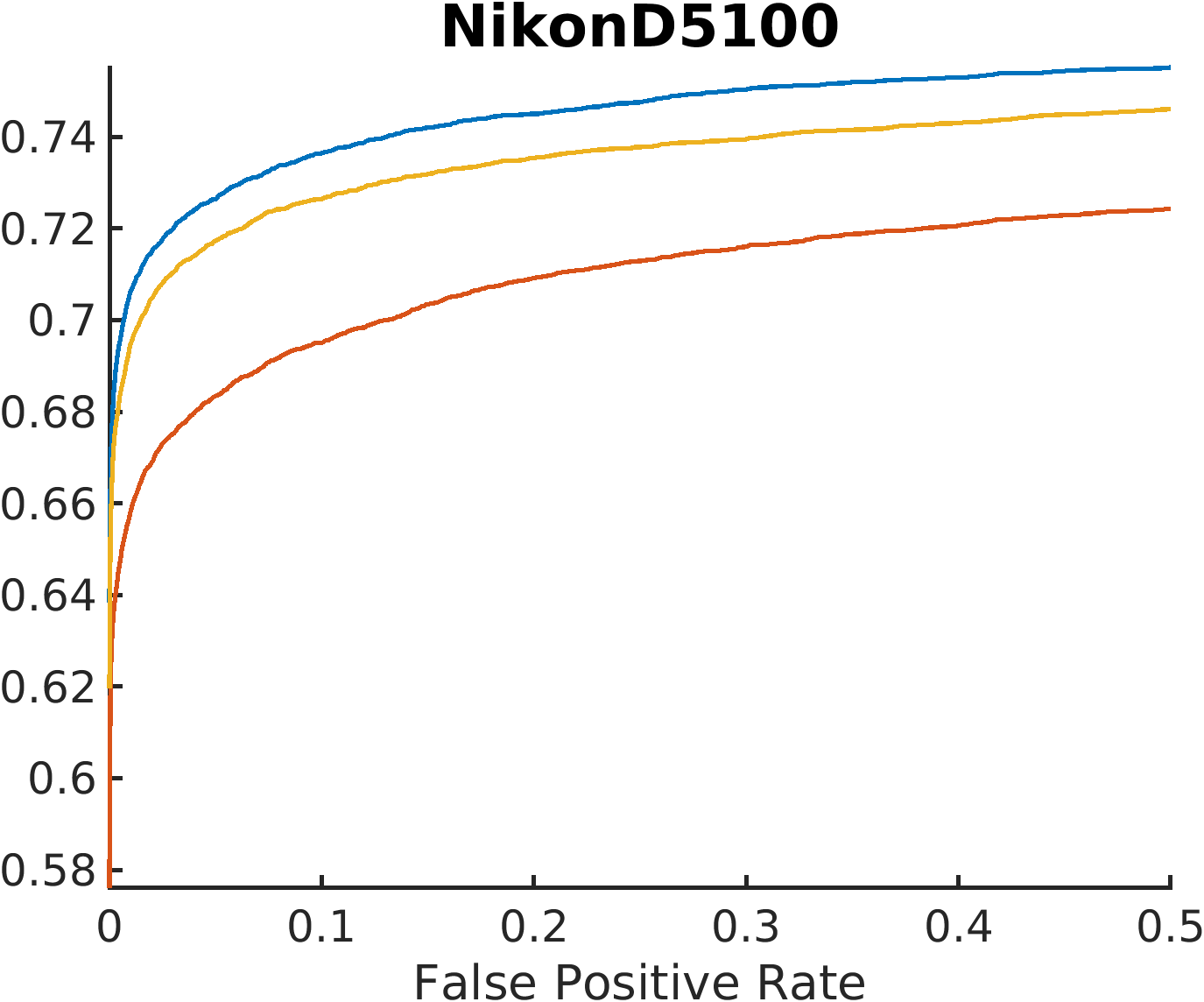"}	
		 \hfill
		\caption{Modified ROC curves in the device identification problem of Sec.~\ref{sec:did}. We followed the conventional setup \cite{ChenFGL2007} (baseline, red) and a modified version including our estimated PRNU emphasis (blue) and a fixed emphasis function (yellow).}
		\label{fig:ROCs}
	\end{figure*}
	\subsection{{Simplified regressogram}}
	\label{sec:simp_reg}
	In the method above, we compute more points than necessary. To simplify it, we recall the estimator of the PRNU \cite{PerezMIC2016}:
	\begin{equation}
		\label{eq:est_k_reg}
		{\hat{\bt k} \doteq \sum\nolimits_{l = 0}^{\nimag-1} \, \x_l \circ \res_l \,\Big /\, \sum\nolimits_{l = 0}^{\nimag-1}\, {(\x_l\circ \x_l)},}
	\end{equation}
	where the division is element-wise. This statistic is accurate if $c \doteq \sum_l\, \x_l \, \circ\, g(\x_l) / \sum_{l}\,\x_l^2$ is independent of the denoised images. Then, we can use a {$1$D} regressogram on the product of the residuals $\res_i$ with $\hat{\bt k}$, as we did in Section \ref{sec:reg} but considering only one covariate. 
	
	To ensure $c$ remains independent of the denoised images, we can refine the estimate iteratively: let $g_1$ be the result of the estimation above. Then, for $m = 1, 2, \hdots$, we evaluate
	\begin{equation*}
		{\hat{\bt k}_{{m+1}} \doteq {\sum\nolimits_{l = 0}^{\nimag-1} g_{{m}}´(\x_l) \circ \res_l}\, \Big/\, {\sum\nolimits_{l = 0}^{\nimag-1} {(g_{{m}}(\x_l)\circ g_{{m}}(\x_l)})},}
	\end{equation*} 
	repeating the process above to estimate $g_{m+1}(x)$. We expect $\sum_l\, g_{{m}}(\x_l) \circ g(\x_l) / \sum_{l}\,g_{{m}}^2(\x_l)$ to be closer to the identity as $m$ grows. Often, two iterations provide an accurate estimate.
	\subsection{Estimating the transfer curve}
	\label{sec:CRF}
	In this section, we approximate the transfer curve from the PRNU emphasis. Let us start from the definition of $g(z)$, adding a constant $a$ to account for the boundary conditions:
	\begin{equation}
		a\, g(z) = z \, h'(z) \implies a\, g(z) \, / \, z = h'(z).
	\end{equation}
	This is a first order differential equation. We know $g(h(z))$, because $h(z)$ is the output of the denoiser. Then, applying the Fundamental Theorem of Calculus and arranging,
	\begin{equation}
		\int_{\epsilon}^{u}\, a\, \frac{g(h(z))}{h(z)} \diff{h(z)} = \int_{\epsilon}^{u}\, h'(h(z)) \diff{h(z)}.
	\end{equation}
	The expression above is valid if $\epsilon \leq h(z) \leq u$. Now,
	\begin{equation}
		a\,G(u)  \doteq  a\, \int_{\epsilon}^{u} \, \frac{g(h(x))}{h(x)} \diff{h(x)} = h(u) - h(\epsilon).
	\end{equation}
	We can evaluate this integral numerically. Our definition of transfer curve requires $h(0) = 0$, thus $\epsilon = 0$. From $h(1) = 1$, we obtain $a = 1\, \big /\, G(1)$. To avoid computational instability in the division by $h(z)$, we set $\epsilon$ to a small value and fix $h(\epsilon) = \epsilon$. We can use $a$ to normalize the PRNU emphasis.
	\section{Empirical results}
	\label{sec:exper}
	In the following, we consider a database of {$15$} cameras\footnote{List of devices. Dresden \cite{GloeB2010}: Two Nikon(N)D70, two ND70S, two ND200. RAISE \cite{DangPCB2015}: One ND7000. Private: Three Canon(C)D1100, one CD600, one ND3200, one ND3000, one ND5100 and one ND60.}. For each of these cameras, we gather $40$ residuals from the denoiser proposed by Mıhçak \etal \cite{MihcakKRM1999}. We work with uncompressed (TIFF) images. We denoise each color channel independently and convert the result to gray-scale. We also apply zero-meaning and Wiener filtering \cite{ChenFGL2008}.
	\subsubsection{PRNU-emphasis and {transfer curve}}
	We estimate the PRNU emphasis using the regressogram and the simplified regressogram with $m = 1$ and $2$, all of them with $32$ bins. We sample $10$ residuals randomly from the database. We follow a leave-one-out strategy to obtain $10$ products of the denoising residuals with the estimated fingerprint $\bt w_l \circ \, \hat{\bt k}_m$, which are used as the input to the simplified regressogram. To compute the confidence bands, we repeat this process $50$ times.
	
	The results show that the PRNU emphasis differs from the linear function assumed by the model in \eqref{eq:chenetal} (Fig.~\ref{fig:regress_real}). Also, the difference between the regressogram and the simplified regressogram after two iterations is negligible.
	
	We compute the corresponding transfer curves following the procedure in Sec. \ref{sec:CRF} (Fig.~\ref{fig:CRFs}). As assumed in Sec. \ref{sec:emph_th}, they are monotonically increasing functions.
	\subsubsection{Device identification}
	\label{sec:did}
	We address a device identification problem with cropped images. Our statistic for detection is the PCE \cite{GoljanF2008}. Whenever we fail to align the residual under $H_1$, we ensure that the sample is labeled as $H_0$. Under $H_0$, we take the maximum of the PCE over all possible alignments.  We summarize our results in a modified ROC curve whose TPR is upper bounded by the probability of finding the correct cropping point under $H_1$ (Fig.~\ref{fig:ROCs}).  
	
	For each device, we use $10$ residuals to estimate $g(x)$ via the simplified regressogram ($m = 2$) with $32$ bins. We divide the other $30$ residuals in half for the training and testing splits. 
	
	In the training stage, we select a square of $2000\times 2000$ pixels to have the same size for all images regardless of the camera. The origin of the square is random albeit known. We compute the estimator of $\hat{\bt k}$ in \eqref{eq:est_k_reg} (baseline), the same estimator using $g(\bt x)$ instead of $\bt x$, and a simplified version using an inverse parabola as ``fixed'' emphasis, $g_{\it f}(x) = -x^2 + 255\, x$.

	In the testing stage, we select first the square of $2000\times 2000$ pixels. Under $H_1$, we set the origin as in the training stage; under $H_0$, the size of the images varies, so we select a random origin. Then, we crop this square to patches of size $512\times 512$ pixels with a random cropping point, which is the quantity to estimate in the alignment process. For $H_1$, we use the remaining $15$ residuals; for $H_0$, we mix $100$ images from the other devices.
	
	Finally, we shuffle the $40$ residuals used on each stage randomly $10$ times. Regarding the squares of $2000\times 2000$ pixels, we change the origin randomly $10$ times, while we consider $10$ different cropping points in the testing part.
	
	Fixing the FPR to $0.01$, the TPR (bounded by the probability of finding the correct cropping point) improves by a $4.93\,\%$ on average with the estimated emphasis and by a $4.16\,\%$ on average for the fixed emphasis for the 15 cameras we tested.  
	\section{Conclusion}
	We have found that the response of the camera, composed with the optical and digital preprocessing pipelines, induces an emphasis function in the denoising residuals. An estimation method is proposed to retrieve this emphasis function. Results have shown that this curve deviates from the linear function assumed by the multiplicative model. Including the PRNU emphasis in a device identification scenario increases the average TPR for the cameras we tested. Finally, using an inverse parabola as fixed emphasis provides competitive results too. Future work will incorporate the PRNU emphasis in estimation and detection problems.
	\appendix
	\label{app:recover_chen}
	This result relates the models in \eqref{eq:chenetal} and \eqref{eq:newmod}.
	\begin{proposition}
		Let $x \doteq h(z)$. Then, the output of the camera response function can be written as
		\begin{equation}
			y = x \, (1+ c \, k) + \bigo{\pow\, {k^2}},
		\end{equation}
		for some real constant $c$ if and only if $h(z) = c_1 z^\gamma$ for some real positive constants $c_1$ and $\gamma$.
	\end{proposition}
	To prove the "if" part, replace $h(z) = c_1 z^\gamma$ in \eqref{eq:newmod}. The "only if" part follows from solving the following equation
	\begin{equation*}
		h(z)( 1 + c \, k) = h(z)+ z h'(z) c_1 \, k \implies c_2 = z \,h'(z) / h(z).
	\end{equation*}
	Rewriting as $h'(z) / h(z) = c_2/z$ and integrating, we obtain $\log(h(z)) = c_2\log(z)+c_3$; thus, $h(z) = e^{c_3}\, z^{c_2}$.
	\bibliographystyle{IEEEtran}
	\bibliography{arXiv_gx.bib}
\end{document}